\input{hyperlatex}
\documentstyle[preprint,eqsecnum,aps]{revtex}

\newcommand{\f}[2]{\frac{#1}{#2}}
\newcommand{\s}[1]{\sqrt{#1}}

\begin{document}

\preprint{\vbox{
\hbox{UCSD/PTH 96--06}
}}
\title{Heavy Baryons in SU(2) $\times$ SU(6)}
\author{Richard F. Lebed\footnote{rlebed@ucsd.edu}}
\address{Department of Physics, University of California at
San Diego, La Jolla, CA 92093}
\date{March 1996}
\maketitle
\begin{abstract}
	The spectrum of baryons containing heavy quarks of one flavor
is described in terms of representations of the group SU(2) $\times$
SU(6), where the two factor groups refer to spin rotations of the
heavy quarks and spin-flavor rotations of the light quarks,
respectively.  This symmetry has a natural interpretation in the heavy
quark limit.  We exhibit the decomposition of baryon mass operators
under this symmetry and compare to experimental results.  We discuss
the relation of this analysis to that of large-$N_c$ QCD as well as
four-flavor SU(8), and indicate the generalization of this work to
other properties of heavy baryons.
\end{abstract}

\pacs{11.30.Hv, 12.40.Yx, 14.20.Lq, 14.20.Mr}


\narrowtext

\section{Introduction}

	Recent experiments continue rapidly to expand our knowledge of
the properties of heavy-quark hadrons.  For example, the past few
years have seen evidence for the first observations of numerous
ground-state charmed and bottom baryons, including both charmed and
bottom cascades and bottom $\Sigma$'s.  One important task of current
theoretical efforts is to perform a critical analysis of whether we
understand the information this new data is providing.  A natural
starting point is to develop an understanding of the mass spectra of
heavy-quark hadrons.  In this paper we propose a symmetry group for
the heavy baryons and explore its mathematical and phenomenological
consequences.

	The symmetry paradigm we adopt is the group SU(2) $\times$
SU(6), where the first factor refers to the spin of the heavy quark $Q
= c$ or $b$, and the second factor is the spin-flavor symmetry of the
three light quarks $u$, $d$, and $s$.  By organizing the
representations (hereafter {\it reps\/}) of the symmetry group in this
way, we recognize the fundamental phenomenological difference between
heavy and light quarks.  In particular, we appear to inhabit a world
in which one may accurately calculate physical quantities by
performing expansions about the massless quark limit for light quarks
(chiral perturbation theory) and about the infinite mass limit for
heavy quarks (heavy quark effective theory).  The symmetry SU(6) for
baryons has a long and illustrious history, and appears to accurately
model reality in describing features like the closeness of the octet
and decuplet of light baryons, the magnetic moment ratio $\mu_p /
\mu_n \approx -3/2$, and the axial current coefficient ratio $F/D
\approx 2/3$.  The decomposition of light baryon bilinear
operators in SU(6), analogous to the analysis performed here for
heavy-quark baryons, appears in Ref.~\cite{I}.  The use of SU(6)
$\times$ O(3) to describe just the light quarks (including orbital
angular momentum) in heavy baryons has recently been advocated by
K\"{o}rner\cite{Korn} to increase predictive power beyond heavy quark
effective theory.

	The ground-state charmed baryons (and by inference, the bottom
baryons) appear to fall into multiplets determined by the SU(4) flavor
symmetry of the $u$, $d$, $s$, and $c$ quarks\cite{PDG}.  How can this
be when the charm quark is so much heavier than the other quarks?  In
fact, the same spectrum arises from the much milder assumption of
approximate SU(3) symmetry for light quarks and the assertion that the
color wavefunction for each baryon is completely antisymmetric in all
of the quark indices.  Then the (spin $\times$ flavor $\times$ space)
wavefunction must be completely symmetric under exchange of quark
indices; since the ground-state baryons are assumed to have no
internal orbital angular momentum, the space wavefunction is
symmetric.  The two possible spins from three quarks are 1/2 (mixed
symmetry) and 3/2 (completely symmetric), and working out the
corresponding SU(3) reps for 3, 2, 1, and 0 light quarks that leave
the product of spin and flavor wavefunctions symmetric gives the
multiplet structure indicated in Ref.~\cite{PDG}.  Unlike for SU(4) or
its corresponding spin-flavor group SU(8), the levels of the
multiplets, each of which has a different number of heavy quarks,
belong to different reps in this construction and thus are {\it a
priori\/} unrelated.

	In principle, any symmetry may be used to describe the heavy
baryons as long as it contains operators in all allowed reps that
contribute to physical quantities; the relation between two such
symmetries is simply a basis transformation.  For example,
Jenkins\cite{Jenk} considers the heavy baryons in the large-$N_c$ QCD
expansion.  It is only in assuming that operators in certain reps give
smaller contributions than others to physical quantities, that one
obtains relations between observables, and since different symmetries
organize the same space of operators in different manners, distinct
predictions arise.

	In Sec.\ 2 we review current experimental knowledge of heavy
baryon masses.  Section 3 presents the tensor formalism for baryons in
SU(2) $\times$ SU(6) and explains how the relevant Clebsch--Gordan
coefficients are obtained.  In Sec.\ 4 we describe the
phenomenological application of the symmetry and compare the results
to experiment.  Section 5 compares the consequences of this symmetry
to spin-flavor SU(8), large-$N_c$ QCD, and other recent work.  Section
6 briefly discusses other applications of the formalism and concludes.
The Clebsch--Gordan coefficients and mass operator decompositions
appear in the Appendix.

\section{Status of Experimental Results}

	Let us briefly review the state of experimental measurements,
both to indicate the level of completeness of the multiplets and to
fix notation for degenerate states.  See \cite{PDG} for a geometrical
picture of the weight space.  The $Q=0$ levels of the multiplets are
the well-known SU(3) {\bf 8} for spin-1/2 and {\bf 10} for spin-3/2,
whereas none of the $Q=2$ or $Q=3$ baryons have yet been observed.
Signals for almost all of the $Q=1$ charmed baryons have been seen,
excepting the $\Omega^*_c$ and some of the distinct isospin states.
For the $Q=1$ bottom baryons, experimental uncertainties on
$\Lambda_b$ mass measurements are rapidly decreasing, preliminary
values for the $\Sigma_b$ and $\Sigma^*_b$ masses exist, and evidence
for the $\Xi_b$ baryon has been presented\cite{xilep}, although mass
measurements have not yet appeared.  The data is summarized in
Table~\ref{mass}.

	Note that two distinct $\Xi_Q$ baryons occupy the same sites
in the multiplets, and thus mixing terms between them occur.  We use
the notation recently favored by experimental groups, that
$\Xi^\prime_Q$ and $\Xi_Q$ respectively indicate the sextet and
antitriplet states of SU(3).  In sextets (antitriplets), the two light
quarks are symmetric (antisymmetric) under exchange of indices, and
thus are in a relative spin-1 (spin-0) state.  One expects
$\Xi^\prime_Q > \Xi_Q$ from a simple quark model-inspired analysis of
the spin-spin coupling in each baryon, in which aligned spins repel
and anti-aligned spins attract.  Another notation for these
particles\cite{Sav} is to use a subscript 1,2 to indicate the light
quarks in a $\bar{\bf 3}$ ({\bf 6}).

\section{Tensor Analysis in SU(2) $\times$ SU(6)}

	The analysis presented here of SU(2) $\times$ SU(6) group
theory by means of tensors closely parallels that in Ref.~\cite{I}
(hereafter called (I)).  We begin by defining appropriately
symmetrized tensors to keep track of quark flavor and spin indices.
In essence, this construction encapsulates all of the
group-theoretical information and thus provides a route for obtaining
all the relevant Clebsch--Gordan coefficients.  Although the
manipulations that follow apply to bottom as well as charmed baryons,
we present the results for the latter; the corresponding results for
the $b$-, $bb$-, and $bbb$-baryons are obtained by subtracting one,
two, and three units from the charge superscript of $c$-, $cc$-, and
$ccc$-baryons, respectively.

	We begin by noting that light quarks in SU(6) transform
according to the fundamental {\bf 6} rep.  For ground-state baryons
with one heavy quark, the two light quarks are completely symmetric
with respect to exchange of (spin $\times$ flavor) indices, owing to
the antisymmetry of the baryon wavefunction under color.  The light
diquark is then in the symmetric SU(6) rep from ({\bf 6} $\otimes$
{\bf 6}), which is the {\bf 21}.  For ground-state baryons with two
heavy quarks, the light quark is necessarily in a {\bf 6}.  The
analogous wavefunction for the $Q=3$ baryon $\Omega^{*+++}_{ccc}$ is
trivial.  In (I) we considered baryons with no heavy quarks, for which
the completely symmetric SU(6) rep is the {\bf 56}.

	The tensor completely symmetric under the exchange of paired
spin and flavor indices for the light diquark system (a {\bf 21} rep
of SU(6)), which leads to the $Q=1$ baryon tensor, may be represented
by
\begin{equation} \label{the21}
{\cal B}^{ai,bj} \equiv \chi^{ij} B^{ab} + \frac{1}{2} \epsilon^{ij}
\epsilon^{abc} B_c ,
\end{equation}
where $\epsilon^{ij}$ and $\epsilon^{abc}$ are Levi-Civita tensors,
$\chi^{ij}$ represents the symmetric spin-one tensor for the
light diquark,
\begin{equation} \label{chi2}
\chi^{ij} = \left( \begin{array}{rr} | 1, +1> & \frac{1}{\sqrt{2}}
| 1, 0> \\ \frac{1}{\sqrt{2}} | 1, 0> & |1, -1> \end{array} \right),
\end{equation}
and the $SU(3)$ baryon tensors $B$ are constructed as follows:  The
{\bf 6} is assigned the entries of the symmetric tensor $B^{ab}$ by
\begin{equation}
\begin{array}{rcrrcrrcr}
B^{11} & = & \Sigma^{++}_c, \, & B^{12} & = & \frac{1}{\sqrt{2}}
\Sigma^+_c, \, & B^{22} & = & \Sigma^0_c, \\
B^{13} & = & \frac{1}{\sqrt{2}} \Xi^{\prime +}_c, \, & B^{23} & = &
\frac{1}{\sqrt{2}} \Xi^{\prime 0}_c, \, & B^{33} & = & \Omega^0,
\end{array}
\end{equation}
and the $\bar{\bf 3}$ is assigned the entries of the tensor $B_c$, in
a particular phase convention\cite{phase}, according to
\begin{equation}
B_1 = \Xi^0_c, \, \, B_2 = -\Xi^+_c, \, \, B_3 = \Lambda^+_c .
\end{equation}
Because these tensors describe diquarks rather than the full heavy
baryon, the identity of the baryon only becomes fixed when the spin of
the heavy quark is included.  In particular, the entries of the {\bf
6} should be taken to represent either spin-1/2 or spin-3/2 baryons,
depending upon the spin of the heavy quark.  Then the full baryon
tensor wavefunction, including the heavy quark spinor $\chi^k$ with
$\chi^{1,2} = \, \uparrow, \downarrow$, is
\begin{equation}
{\cal B}^{ai,bj;k} \equiv {\cal B}^{ai,bj} \chi^k.
\end{equation}
Lastly, the factor 1/2 in Eq.~(\ref{the21}) is determined by the
singlet normalization that $\overline{\cal B}^{ai,bj;k} {\cal
B}^{ai,bj;k}$ produces all bilinears for each baryon field in a given
spin state appearing only with its conjugate, and each with
coefficient unity.

	The tensor rep for baryons consisting of two heavy quarks and
one light quark is much simpler.  The light quark piece ({\bf 6} of
SU(6)) is just
\begin{equation}
{\cal B}^{ai} = B^a \chi^i ,
\end{equation}
where $\chi^i$ is its spin, with $\chi^{1,2} = \, \uparrow,
\downarrow$, and the {\bf 3} of SU(3) is assigned the entries of the
tensor $B^a$ according to
\begin{equation}
B^1 = \Xi^{++}_{cc}, \, \, B^2 = \Xi^+_{cc}, \, \, B^3 = \Omega^+_{cc}
.
\end{equation}
As before, these components may refer to either spin-1/2 or spin-3/2
baryons, depending upon the spin state of the two heavy quarks.  The
full baryon tensor is then
\begin{equation}
{\cal B}^{ai;jk} = {\cal B}^{ai} \chi^{jk} ,
\end{equation}
where $\chi^{jk}$ is the (symmetric) spin tensor for the two heavy
quarks and has the same form as Eq.~(\ref{chi2}).

	The SU(2) $\times$ SU(6) decomposition using these tensors now
follows from the same methods as in (I): Baryon mass terms are
bilinears with $J_3 = I_3 = Y = 0$ and total spin $J=0$, so the SU(6)
{\bf 1} and {\bf 35} combinations are obtained by computing
the bilinear expressions
\begin{equation}
\overline{\cal B}^{ai,bj;k} {\cal T}_{bj}{}^{c \ell} {\cal J}_k{}^m
{\cal B}_{ai,c \ell;m} \, ,
\end{equation}
for $Q=1$, and
\begin{equation}
\overline{\cal B}^{ai;jk} {\cal T}_{ai}{}^{b \ell} {\cal J}_k{}^m
{\cal B}_{b \ell;jm} \, ,
\end{equation}
for $Q=2$, where ${\cal T}$ are light-quark spin-flavor generators and
${\cal J}$ are heavy-quark spin generators.  It is enough to
compute explicitly the tensor for ${\cal T} \otimes {\cal J} = 1
\otimes 1$, $ T^3 \otimes 1$, $T^8 \otimes 1$, $J_3 \otimes j_3$,
$(T^3 J_3) \otimes j_3$, and $(T^8 J_3) \otimes j_3$, because these
give the SU(6) {\bf 1} and {\bf 35}, and the {\bf 405} for $Q=1$ may
be found by orthogonality.

	In the Appendix we begin by computing the combinations of
bilinears (``chiral coefficients'') transforming under particular
SU(3) and isospin reps, and combine them into SU(6) chiral
coefficients by means of the tensor methods just outlined.  The only
complication is that one must take care to project out the appropriate
components of heavy quark spin to obtain baryons with the desired
total spin.  Chiral coefficients not involving a spin flip of the
heavy quark (heavy and light quark bilinears each with $j=0$) are
labeled $X$; those with a spin flip (heavy and light quark bilinears
each with $j=1$), and therefore suppressed in the infinite quark mass
limit, are labeled $Y$.  Similarly, the mass combinations associated
with each chiral coefficient are labeled with the corresponding
lower-case letter $x$ or $y$.

	There are 18 distinct mass combinations for the $Q=1$ baryons
and 6 for $Q=2$, and these numbers are borne out by the mass
combinations listed in Eqs.~(\ref{q1rel}--\ref{q2rel}).  The former
number is obtained by noticing that, in addition to two sextets and
one antitriplet, there is a mixing parameter between each member of
the $\bar {\bf 3}$ and the state in the spin-1/2 {\bf 6} with the same
weight.

\section{Phenomenological Consequences}

	In order to estimate the expected sizes of mass combination
coefficients $x$ and $y$ listed in the Appendix, one must make some
assumptions regarding the pattern of symmetry breaking; this analysis
is similar to that in Ref.~\cite{JL}.  First note that all
combinations in Eqs.~(\ref{q1rel}--\ref{q2rel}) have zero net baryon
and charm number except for the overall singlet term $x^{1,0}_1$, and
therefore vanish in the SU(2) $\times$ SU(6) limit.  Then the amount
by which each combination deviates from zero is determined by the
finding its overall scale and factors associated with symmetry
breaking.  To accomplish this, the combination is set to zero in the
form $lhs = rhs$, where $lhs$ and $rhs$ are combinations of baryon
masses with positive coefficients.  Dividing by one-half of the sum of
the numerical coefficients on either side ($\equiv k/2$) gives a
scale-independent result, and the magnitude of the combination is set
by the typical {\it uncharmed\/} baryon mass $\Lambda_\chi \approx$ 1
GeV (uncharmed because the full combination has net zero charm
number).  Before including explicit symmetry breaking, the combination
naively satisfies $|x^{R,I}_N| \raisebox{-0.5ex}{$\stackrel{\textstyle
<}{\sim}$} k/2$, although in this expression we neglect an unknown
coefficient of order unity that may make precise value of the
combination correspondingly larger or smaller than this estimate.  The
factor of 1/2 places the uncertainty in the combination symmetrically
about zero.

	For the light diquark, symmetry breaking appears in the
adjoint {\bf 35} rep of SU(6) in the form of relative spin flips,
SU(3) breaking, and $I=1$ isospin breaking, the latter two being
respectively parametrized by $\epsilon$ and $\epsilon^\prime$.  $I=2$
isospin breaking, parametrized by $\epsilon^{\prime\prime}$, first
appears in the {\bf 405}, with its leading contribution arising from
electromagnetic effects.  Since spin operators may only appear in
pairs in mass bilinears, light-quark spin flips first appear in the
{\bf 405} and are parametrized by $\delta$, and spin flips involving
the heavy quark are parametrized by a factor $\theta \approx
\Lambda_{\rm QCD} / m_c$.  For example, the combination
$x^{8,0}_{405}$ has an anticipated size of $12 \Lambda_{\chi} \delta
\epsilon$, with the scale from the arguments of the previous
paragraph, and two factors of SU(6) breaking because the {\bf 405}
requires a product of two {\bf 35}'s, which must include SU(3)
breaking because the SU(3) content is $I=0$ octet.  The anticipated
magnitudes of the combinations are listed in Table~\ref{comb}.

	The set of baryon mass differences is thus reduced to the
scale $\Lambda_{\chi}$ and the dimensionless parameters $\delta$,
$\epsilon$, $\theta$, $\epsilon^\prime$, and
$\epsilon^{\prime\prime}$.  We estimate $\Lambda_{\chi} \approx$ 1
GeV, $\delta \approx \epsilon \approx 0.3$, $\theta \approx 0.2$,
$\epsilon^\prime \approx 0.005$, and $\epsilon^{\prime\prime} \approx
0.001$.  This value for $\delta$ comes from the observation that the
spin-flip operator explains the fractional difference of light octet
and decuplet baryons, $\epsilon$ and $\epsilon^\prime$ respectively
arise from $m_s /\Lambda_{\chi}$ and $(m_u - m_d)/\Lambda_{\chi}$
effects, and $\epsilon^{\prime\prime}$ arises from noting that $I=2$
effects occur in electromagnetic terms of $O(\alpha
\Lambda_{\chi}/4\pi)$.  These values should be taken as indicative
rather than definitive, but the basic pattern should remain.

	From Table~\ref{comb} it is clear that one should focus upon
the combinations associated with the largest reps, where the estimated
magnitudes are smallest.  Hyperfine ($y$) combinations are also highly
suppressed.  One caveat is that when $x$'s or $y$'s are combined, the
numerical coefficient and suppressions like those in Table~\ref{comb}
must be recomputed for the particular combination.  For the moment,
let us assume that mass mixings are negligible, so that the observed
states are SU(3) eigenstates, although we will see that this may not
be true for $\Xi^\prime_c$ and $\Xi_c$.

	We begin by considering a combination to which Savage~\cite{Sav}
computed SU(3) chiral loop corrections, $y^{27,0}_{405}$.  Taking the
experimental numbers in Table~\ref{mass} at face value, we predict
\begin{equation}
\Omega^*_c = 2790 \pm 31 (\pm 36) \, {\rm  MeV},
\end{equation}
where the first error is from experimental uncertainties and the
second follows from analysis as in Table~\ref{comb}.  As for the
poorly-known $\Xi^\prime_c$ mass, we may either check the measured
value as it appears in $\f{1}{5} (y^{8,0}_{35} + 2y^{27,0}_{405}) \,
\raisebox{-0.5ex}{$\stackrel{\textstyle <}{\sim}$}
\Lambda_\chi \theta \epsilon \approx 60$ MeV:
\begin{equation}
(\Sigma^*_c - \Sigma_c) - (\Xi^*_c - \Xi^\prime_c) = -7 \pm 17 \, {\rm
MeV},
\end{equation}
which is certainly consistent with our estimates, or use it to predict
the $\Xi^\prime_c$ mass:
\begin{equation}
\Xi^\prime_c = 2567 \pm 7 \, (\pm 60) \, {\rm MeV} .
\end{equation}
From this example we see that, as one is forced to employ
less-suppressed combinations (here $y^{8,0}_{35}$), the theoretical
uncertainty becomes too large to make very useful predictions for
individual masses.  Particular models can do much better, of course,
because in such cases the dynamical assumptions are much more
restrictive.  For the remainder of the $I=0$ combinations, we seek to
demonstrate that the values associated with each combination in
Table~\ref{comb} are reasonable by eliminating from each the unknown
$\Omega^*_c$ and poorly-known $\Xi^\prime_c$ masses.  These results
are presented in Table~\ref{compare}.  The first line in the table
exhibits the size of the overall singlet of the $C=1$ multiplet, ${\rm
M}_c = 2558 \pm 2 ({\rm expt.}) \pm 3 ({\rm theor.})$ MeV, whereas
comparing the second and fifth columns in the succeeding lines
indicates that the symmetry with our choice of expansion parameters is
reliable, although the experimental smallness of the final $I=0$
combination is notable.

	The known $I=1$ mass combinations are $\Sigma_{c1} \equiv
(\Sigma^{++}_c - \Sigma^0_c)$, $\Xi_{c1} \equiv (\Xi^+_c - \Xi^0_c)$,
and $\Xi^*_{c1} \equiv (\Xi^{*+}_c - \Xi^{*0}_c)$, while
$\Sigma^*_{c1} \equiv (\Sigma^{*++}_c - \Sigma^{*0}_c)$ and
$\Xi^\prime_{c1} \equiv (\Xi^{\prime +}_c - \Xi^{\prime 0}_c)$ are
unknown.  The predictions for the latter are
\begin{eqnarray}
\Sigma^*_{c1} & = & \Sigma_{c1} + \f{1}{5} (2y^{8,1}_{35} +
y^{27,1}_{405}) \nonumber \\
& = & 0.7 \pm 0.4 \, (\pm 1) \, {\rm MeV}, \\
\Xi^\prime_{c1} & = & \Xi_{c1} - \f{1}{5} (y^{8,1}_{35} -
2y^{27,1}_{405}) \nonumber \\
& = & -5.2 \pm 2.2 \, (\pm 1) \, {\rm MeV},
\end{eqnarray}
where in both cases the scale of the theoretical uncertainty is
$\Lambda_\chi \theta \epsilon^\prime \approx$ 1 MeV.  The results from
eliminating $\Sigma^*_{c1}$ and $\Xi^\prime_{c1}$ from the other
combinations appear in Table~\ref{compare}; the experimental smallness
of the first $I=1$ combination in the table and the largeness of the
second compared to estimates may be related to the problem of
$\Xi^\prime$-$\Xi_c$ mixing, since both contain $\Xi_{c1}$, whereas
the combination of the two eliminating $\Xi_{c1}$ is of expected size.

	The analysis for $I=2$ mass combinations predicts
\begin{eqnarray}
(\Sigma^{*++}_c - 2\Sigma^{*+}_c + \Sigma^{*0}_c) & = & (\Sigma^{++}_c
- 2\Sigma^+_c + \Sigma^0_c) + y^{27,2}_{405} \nonumber \\
& = & -2.1 \pm 1.3 \, (\pm 0.4) \, {\rm MeV} ,
\end{eqnarray}
while eliminating the $\Sigma^*_c$ combination gives the final line in
Table~\ref{compare}.

	Finally, we consider $\Sigma^+_c$-$\Lambda^+_c$ and
$\Xi^\prime$-$\Xi$ mixings.  Performing the diagonalization of a 2
$\times$ 2 mass matrix with diagonal entries $\mu_{1,2}$ and mixing
$\nu$ gives the eigenvalues
\begin{equation}
\mu^{\rm phys}_{1,2} = \f{\mu_1 + \mu_2}{2} \pm \s{\left(\f{\mu_1 -
\mu_2}{2} \right)^2 + \nu^2} .
\end{equation}
When the mixing $\nu$ is pure $I=1$, it is not only suppressed by
$\epsilon^\prime$, but much more so through the Pythagorean sum.  Thus
the mixing $\gamma$ as defined in the Appendix has very little effect
on the masses of $\Sigma^+_c$ or $\Lambda^+_c$, as was the case for
$\beta$ with $\Sigma^0$ and $\Lambda$ in (I).  On the other hand,
$\Xi^\prime_c$-$\Xi_c$ has both $I=1$ and $I=0$ mixings, proportional
to $(\delta_+ \pm \delta_0)$.  The $I=0$ mixing parameters may be as
large as 60 MeV, so this mixing may contribute tens of MeV to the
splitting between $\Xi^\prime_c$ and $\Xi_c$.  In fact, since one-half
the observed splitting between the physical $\Xi^\prime_c$ and $\Xi_c$
is about 45 MeV, it is quite possible that the pure {\bf 6}
$\Xi^\prime_c$ and pure $\bar {\bf 3}$ $\Xi_c$ could be degenerate in
mass.  Certainly a more detailed analysis of $\Xi_c$ masses must
either take this mixing into account or else explain why it is
suppressed below its natural size.

\section{Other Schemes for Heavy Baryon Masses}
\subsection{Relationship to SU(8)}

	The discovery of charm in 1974 prompted much analysis of
hadronic properties in terms of an assumed four-flavor symmetry SU(4)
among the quark flavors ($u,d,s,c$) then known.  SU(4) was naturally
extended to the spin-flavor symmetry SU(8), in analogy with the
spin-flavor symmetry SU(6) among the three light quark flavors studied
extensively in the 1960s.  However, the four-flavor symmetries are
badly broken because the charm quark is not only much heavier than
$u$, $d$, $s$, but also the QCD scale $\Lambda_{\rm QCD}$
\raisebox{-0.5ex}{$\stackrel{\textstyle <}{\sim}$} 1 GeV.  Thus one
should not take SU(8) seriously as an accurate description of physical
quantities, but as a mathematical question it is nonetheless
interesting to compare its predictions to those of its subgroup SU(2)
$\times$ SU(6) considered in this work, because the predictions of
SU(8) analysis exist in the literature.

	In analogy to the {\bf 56} rep of SU(6), one
considers the baryons of SU(8) to fill the completely symmetric
rep with Dynkin symbol (3,0,0,0,0,0,0) = {\bf 120}.  We now
recapitulate the arguments in (I) for SU(6).  All group-theoretical
relations between static baryon quantities (masses, magnetic moments,
etc.) are obtained by decomposing the bilinear product
\begin{eqnarray}
\overline{\bf 120} \otimes {\bf 120} & = & {\bf 1} \oplus {\bf 63}
\oplus {\bf 1232} \oplus {\bf 13104} \nonumber \\
& = & (0,0,0,0,0,0,0) \oplus (1,0,0,0,0,0,1) \oplus (2,0,0,0,0,0,2)
\oplus \nonumber \\ & & (3,0,0,0,0,0,3) .
\end{eqnarray}
One now argues that symmetry-breaking effects among the members of the
{\bf 120}, due to quark masses, charges, or relative spin flips, occur
in the adjoint rep {\bf 63}.  Since it possesses one fundamental and
one antifundamental quark index, the adjoint is called a one-body
operator.  But now note that the most general two-body operator
transforms under the reps in {\bf 63} $\otimes$ {\bf 63}, of which the
largest is the {\bf 1232}; thus the largest rep in $\overline{\bf 120}
\otimes {\bf 120}$, the {\bf 13104}, does not appear when three-body
operators are neglected.  The neglect of three-body operators was
precisely the assumption made by Franklin\cite{Frank} when deriving
charmed baryon mass relations, so his results correspond in
group-theoretical terms to the neglect of mass operators transforming
under the {\bf 13104} rep of SU(8).

	To count the number of mass relations provided by this Ansatz,
one decomposes the {\bf 13104} into spin-flavor SU(2) $\times$ SU(4)
reps $(J,R)$ and counts the number of bilinear states among these with
$J=0$ ($J=1$ for magnetic moment relations, and so on), $I_3 = 0$,
$Y=0$, and heavy quark number $Q=0$.  This gives the mass operators in
the {\bf 13104}, which by the Ansatz have vanishing matrix elements
and thus produce relations among the baryon masses.  It turns out that
the decomposition of the {\bf 13104} with $J=0$ gives 32 such states,
of which 4 are removed by the imposition of time reversal invariance
(see Appendix).  Ten of these are the three-body SU(6) {\bf 2695}
relations exhibited in (I), so there must be 18 relations involving
charmed baryons.  Ref.~\cite{Frank} exhibits 14 such relations; when
including the {\bf 6}-$\bar{\bf 3}$ mixing terms $\gamma$ and
$\delta_{+,0}$ defined in the Appendix, we find three more relations
by applying his method to the {\bf 2695} relation involving the
$\Lambda$-$\Sigma^0$ mixing $\beta$ from (I),
\begin{eqnarray}
4\s{3} \gamma & = & (\Sigma^{++}_c - \Sigma^0_c) + (\Xi^{++}_{cc} -
\Xi^+_{cc}) - (\Sigma^{*++}_c - \Sigma^{*0}_c) + (\Xi^{*++}_{cc} -
\Xi^{*+}_{cc}) , \nonumber \\
4\s{3} (\delta_0 - \delta_+) & = & (\Sigma^{++}_c - \Sigma^0_c) +
(\Xi^{++}_{cc} - \Xi^+_{cc}) - (\Sigma^{*++}_c - \Sigma^{*0}_c) +
(\Xi^{*++}_{cc} - \Xi^{*+}_{cc}) , \nonumber \\
4\s{3} (\delta_0 + \delta_+) & = & 2(\Omega_c - \Omega^*_c) +
2(\Omega_{cc} - \Omega^*_{cc}) + (\Sigma^{*0}_c + \Sigma^{*++}_c) -
(\Sigma^{0}_c + \Sigma^{++}_c) \nonumber \\ & & \mbox{}+
(\Xi^{*+}_{cc} + \Xi^{*++}_{cc}) - (\Xi^{+}_{cc} + \Xi^{++}_{cc}) .
\end{eqnarray}
An 18th independent relation derivable from his method but not
appearing in \cite{Frank} is taken to be
\begin{equation}
(\Sigma^{*+} - \Sigma^{*-}) - 2(\Xi^{*0} - \Xi^{*-}) = (\Sigma^{*++}_c
- \Sigma^{*0}_c) - 2(\Xi^{\prime +}_c - \Xi^{\prime 0}_c) .
\end{equation}

	The interesting question is how many of these relations turn
out to be supported in SU(2) $\times$ SU(6).  Recall that SU(2)
$\times$ SU(6) decouples sectors with different values of $Q$, and so
we must look for SU(8) relations in which only baryons with the same
number of heavy quarks occur.  Direct manipulation shows that the only
three-body SU(8) mass relations with a single value of $Q$ are: with
$Q=0$, the ten listed in (I) arising from SU(6) analysis, and with
$Q=1$, using the notation in Eq.~(\ref{q1rel}), the five
relations
\begin{eqnarray}
\s{3} y^{8,0}_{405} = -\f{2}{5} y^{8,0}_{35}, & \hspace{2em} &
y^{27,0}_{405} = 0 , \nonumber \\
\f{1}{\s{3}} y^{8,1}_{405} = \f{1}{5} y^{8,1}_{35}, & \hspace{2em} &
y^{27,1}_{405} = 0, \nonumber \\ & & y^{27,2}_{405} = 0 .
\end{eqnarray}
None of the SU(8) relations can be written solely in terms of $Q=2$ or
$Q=3$ baryons.  Note that all of the $Q=1$ SU(8) relations involve
hyperfine splittings or mixing terms.

\subsection{Relationship to Large-$N_c$}

	The large-$N_c$ QCD expansion for baryons possesses an
approximate contracted spin-flavor symmetry\cite{DJM1}; for $N_f$
light flavors, this symmetry is very similar, although not identical,
to SU(2$N_f$).  The analysis of the light baryon masses in large-$N_c$
was performed in Ref.~\cite{JL}.  The physical baryons known to
experiment are taken to occupy small corners of the large baryon reps
allowed in large-$N_c$, where $J$, $I$, and $Y$ are all $O(1)$, not
$O(N_c)$, and this Ansatz leads to substantial predictive power.  One
proceeds by writing down all allowed 0-, 1-, 2-, and 3-body spin,
flavor, and spin-flavor operators between the quarks, making use of
identities\cite{DJM2} that reduce their number to an easily manageable
set spanning the physical baryon masses.  In general, the higher-body
operators tend to have more $N_c$ suppressions than lower-body
operators.  Some of the operators are accompanied by explicit factors
of 1/$N_c$, whereas others produce factors of $N_c$ when acting upon
the physical baryons.  One may also include explicit factors of SU(3)
and isospin breaking.

	The case of heavy baryons was recently considered by
Jenkins\cite{Jenk}; the main difference from the light baryon case is
that one must also include the heavy quark spin operator in the
analysis.  Note the similarities to this study: In both cases, the
basic one-body operators are spin, flavor, and spin-flavor operators,
and a separate heavy-quark spin operator; and more suppressed
contributions occur with higher-body operators.  The two methods thus
lead to very similar results.  For example, pure 2-body light-quark
operators for $Q=1$ baryons in large-$N_c$ transform the same way as
operators in the SU(6) {\bf 405}, so in both analyses the combinations
$y^{27,I}_{405}$ in Eq.~(\ref{q1rel}) are highly suppressed.  A major
theoretical difference is that the suppression of light-quark
spin-flips in this work is suppressed by the phenomenological
parameter $\delta$, whereas large-$N_c$ includes an explicit factor of
$1/N_c$ with each light-quark spin $J^i$, so this suppression is more
natural in large-$N_c$.

\subsection{Comparison to Other Analyses}

	Here we focus briefly on two recent works on understanding the
heavy baryon spectrum.  First, Zalewska and Zalewski\cite{ZZ} propose
a ``simple-minded'' pattern for heavy baryon masses based on the
following three rules: i) equal spacing between sextet isomultiplets,
ii) equal spacing between corresponding spin-1/2 baryons containing
$c$ and $b$ quarks, and iii) hyperfine splittings inversely
proportional to the heavy quark mass.  In terms of the mass
combinations presented here, these rules correspond to i) $x^{27,0}_N
= y^{27,0}_N = 0$, because the SU(3) singlets do not split masses and
octets alone produce equal-spacing, and $y^{8,0}_{35} = 0$, because
the equal-spacing coefficients cancel; ii) $x^{R,0}_N (c) = x^{R,0}_N
(b)$ for $R \neq 1$, for which the net coefficients of spin-1/2 and
spin-3/2 baryons are separately zero, as well as a nontrivial relation
\begin{equation}
(x^{1,0}_{405} (b) - x^{1,0}_{405} (c)) = +2(y^{1,0}_{35} (b) -
y^{1,0}_{35} (c)) .
\end{equation}
Finally, iii) corresponds to $y^{R,I}_N (c) / y^{R,I}_N (b) =
m_b/m_c$.  To obtain direct relations between $b$ and $c$ mass
combinations here would require the expansion to a symmetry group
including heavy flavor, such as SU(2) $\times$ SU(2) $\times$ SU(6),
or better still, SU(4) $\times$ SU(6), which incorporates the full
heavy quark spin-flavor symmetry.

	The third rule, which arises from heavy-quark flavor symmetry,
is called into question by recent data (see Table~\ref{mass}) for the
ratio $(\Sigma_b^* - \Sigma_b)/(\Sigma_c^* -
\Sigma_c) \approx 0.73 \pm 0.13$, which is rather different from
the expected $m_c/m_b \approx 1/3$.  Based on this observation and the
purported failure of an equal-spacing relation
\begin{equation} \label{falkrel}
\overline{\Sigma}_c - \Lambda_c - \overline{\Xi}^*_c + \Xi_c = 0,
\end{equation}
by as much as 80 MeV (the bar indicates a spin average over
corresponding spin-1/2 and spin-3/2 states), Falk\cite{Falk} proposes
a new identification of the heavy baryons in which the
experimentally-observed states identified as $\Sigma_Q$ and
$\Sigma^*_Q$ are actually $\Sigma^*_Q$ and the orbital excitation
$\Sigma^*_{Q(0)}$, while the true $\Sigma_Q$ decays radiatively and
has not yet been observed.  While the magnitude of the hyperfine ratio
might pose difficulties for heavy quark theory if it persists, the
problem of Eq.~(\ref{falkrel}) is less troubling.  Its magnitude is
given by
\begin{equation}
\f{1}{15} (x^{8,0}_{405} + 2 x^{27,0}_{405})
\raisebox{-0.5ex}{$\stackrel{\textstyle <}{\sim}$}
\Lambda_{\chi} \delta \epsilon \approx 90 \, {\rm MeV} ,
\end{equation} 
which by our estimates can easily accommodate the experimental value.
Moreover, recall that the identification of the physical $\Xi_Q$ and
$\Xi^\prime_Q$ with pure $\bar {\bf 3}$ and {\bf 6} states is
obfuscated by a mixing parameter that {\it a priori\/} leads to mass
shifts as large as tens of MeV.

\section{Conclusions}

	The symmetry group SU(2) $\times$ SU(6) provides a natural
organization for calculating quantities relevant to the heavy baryons.
By construction, it is designed to allow both a heavy-quark expansion
and a light-quark spin-flavor expansion.  In this paper we exhibited
the group-theoretical features of the symmetry by explicitly
constructing the tensor rep of ground-state baryons, and applied this
to the mass spectrum.  We found that the observed particles tend to
fit well into multiplets of this symmetry when natural values for
symmetry-breaking parameters are assumed, although the possibility of
large $\Xi^\prime_c$-$\Xi_c$ mixings may complicate the spectroscopy.

	Other computations, for example comparing decays of different
heavy baryons, may be performed using this symmetry.  By projecting
out bilinears of total spin $J=1,2$, or 3, one may examine the
structure of magnetic dipole, electric quadrupole, or magnetic
octupole moments of the baryons, but it is questionable whether any of
these quantities will be measured in the near future, owing to the
short lifetimes of heavy baryons.  Nevertheless, channeling of
short-lived particles through bent crystals, in which very large
effective magnetic fields are possible, may make such measurements
feasible\cite{E761}.

	Another direction involves a similar tensor analysis for the
orbitally-excited baryons, beginning with the observed
$\Lambda^*_{c1}$ and $\Lambda^*_{c2}$.  In this case, the tensor
(\ref{the21}) is modified to include $\ell = 1$ by the additional
product of a tensor transforming like the spherical harmonic $Y^{1m}$.

	Furthermore, generalization of the symmetry group SU(2)
$\times$ SU(6) to SU(4) $\times$ SU(6), where the SU(4) is the full
spin-flavor group of $b$ and $c$ quarks in heavy quark effective
theory, may yield some interesting results.  The phenomenological
analysis then is supplemented by the additional expansion parameter
$m_c/m_b$.  In particular, one can test whether the experimental ratio
of $\Sigma_b$ to $\Sigma_c$ hyperfine splittings remains much larger
than $m_c/m_b$ when corrections of natural size are taken into
account, analogous to the analysis in Sec.\ 4.

	In its own right, this work provides a framework for comparing
mass calculations through potential models or lattice simulations to
what is expected based on considerations of a physically-motivated
approximate symmetry.  As additional heavy baryons are observed and
their masses are measured with decreasing uncertainties, one will gain
greater insight into the symmetries obeyed --- or not obeyed --- in
the interactions between heavy and light quarks.

\vskip3em
{\it Acknowledgments}
\hfil\break
I wish to thank Elizabeth Jenkins for conversations regarding her
$1/N_c$ analysis of the baryon masses, which inspired this parallel
effort.  This work is supported by the Department of Energy under
contract DOE-FG03-90ER40546.

\appendix
\section*{Decomposition of Mass Operators}

	We begin by defining the ``chiral coefficients'' of SU(3),
which are combinations of baryon bilinears transforming under distinct
reps of SU(3) and isospin.  The relation between the ordinary
bilinears and chiral coefficients is simply a basis change, so they
are related by an orthogonal transformation.  For definiteness of
notation for isospin states, we use charmed baryon labels.  Starting
with $Q=1$ baryons, define
\begin{eqnarray}
{\bf M}_{\bar 6 6} & = & {\cal C}_{\bar 6 6} \, {\bf a}_{\bar 6 6} ,
\nonumber \\
{\bf M}_{\bar 6 \bar 3} & = & {\cal C}_{\bar 6 \bar 3} \, {\bf
b}_{\bar 6 \bar 3} , \nonumber \\
{\bf M}_{36} & = & {\cal C}_{36} \, \overline{\bf b}_{36} , \nonumber \\
{\bf M}_{3 \bar 3} & = & {\cal C}_{3 \bar 3} \, {\bf c}_{3 \bar 3} ,
\end{eqnarray}
where
\begin{eqnarray}
{\bf M}^T_{\bar 6 6} & = & (\overline{\Sigma}^{++}_c \Sigma^{++}_c, \,
\overline{\Sigma}^+_c \Sigma^+_c, \, \overline{\Sigma}^0_c \Sigma^0_c,
\, \overline{\Xi}^{\prime +}_c \Xi^{\prime +}_c,\, 
\overline{\Xi}^{\prime 0}_c \Xi^{\prime0}_c, \, \overline{\Omega}^0_c
\Omega^0_c), \nonumber \\
{\bf M}^T_{\bar 6 \bar 3} & = & (\overline{\Sigma}^+_c \Lambda^+_c, \,
\overline{\Xi}^{\prime +}_c \Xi^+_c, \, \overline{\Xi}^{\prime 0}_c
\Xi^0_c), \nonumber \\
{\bf M}^T_{3 6} & = & {\bf M}^\dagger_{\bar 6 \bar 3}, \nonumber \\
{\bf M}^T_{3 \bar 3} & = & (\overline{\Lambda}^+_c \Lambda^+_c, \,
\overline{\Xi}^+_c \Xi^+_c, \, \overline{\Xi}^0_c \Xi^0_c),
\end{eqnarray}
with an analogous ${\bf M}^T_{\bar 6 6}$ for spin-3/2 states, and
\begin{eqnarray}
{\bf a}^T_{\bar 6 6} & = & (a^1_0, \, a^8_0, \, a^8_1, \, a^{27}_0, \,
a^{27}_1, a^{27}_2) , \nonumber \\
{\bf b}^T_{\bar 6 \bar 3} & = & (b^8_0, \, b^8_1,
\,b^{\overline{10}}_1) , \nonumber \\
\overline{\bf b}^T_{3 6} & = & {\bf b}^\dagger_{\bar 6 \bar 3} ,
\nonumber \\
{\bf c}^T_{3 \bar 3} & = & (c^1_0, \, c^8_0, \, c^8_1) ,
\end{eqnarray}
with upper (lower) indices indicating SU(3) (isospin) reps. Then we
compute\cite{Kaed}
\begin{eqnarray}
{\cal C}_{\bar 6 6} & = &
\left( \begin{array}{cccccc}
+\f{1}{\s{6}}  & +\s{\f{2}{15}}  & +\s{\f{2}{5}}  & +\f{1}{\s{30}} &
+\f{1}{\s{10}} & +\f{1}{\s{6}} \\
+\f{1}{\s{6}}  & +\s{\f{2}{15}}  &             0  & +\f{1}{\s{30}} &
            0  & -\s{\f{2}{3}}  \\
+\f{1}{\s{6}}  & +\s{\f{2}{15}}  & -\s{\f{2}{5}}  & +\f{1}{\s{30}} &
-\f{1}{\s{10}} & +\f{1}{\s{6}} \\
+\f{1}{\s{6}}  & -\f{1}{\s{30}}  & +\f{1}{\s{10}} & -\s{\f{3}{10}} &
-\s{\f{2}{5}}  &             0 \\
+\f{1}{\s{6}}  & -\f{1}{\s{30}}  & -\f{1}{\s{10}} & -\s{\f{3}{10}} &
+\s{\f{2}{5}}  &             0 \\
+\f{1}{\s{6}}  & -2\s{\f{2}{15}} &              0 & +\s{\f{3}{10}} &
             0 &             0
\end{array} \right) ,
\end{eqnarray}
\begin{equation}
{\cal C}_{\bar 6 6} =
\left( \begin{array}{cccccc}
            0 & +\s{\f{2}{3}} & +\f{1}{\s{3}} \\
+\f{1}{\s{2}} & +\f{1}{\s{6}} & -\f{1}{\s{3}} \\
+\f{1}{\s{2}} & -\f{1}{\s{6}} & +\f{1}{\s{3}}
\end{array} \right) ,
\end{equation}
\begin{equation}
{\cal C}_{3 \bar 3}  = 
\left( \begin{array}{cccccc}
+\f{1}{\s{3}} & +\s{\f{2}{3}} &             0 \\
+\f{1}{\s{3}} & -\f{1}{\s{6}} & +\f{1}{\s{2}} \\
+\f{1}{\s{3}} & -\f{1}{\s{6}} & -\f{1}{\s{2}}
\end{array} \right) .
\end{equation}
For the states with $Q=2$,
\begin{equation}
{\bf M}_{\bar 3 3} = {\cal C}_{\bar 3 3} \, {\bf d}_{\bar 3 3} ,
\end{equation}
where we may choose the phase convention
\begin{equation}
{\cal C}_{\bar 3 3} = {\cal C}_{3 \bar 3} ,
\end{equation}
with
\begin{equation}
{\bf M}^T_{\bar 3 3} = (\overline{\Omega}^+_{cc} \Omega^+_{cc}, \,
\overline{\Xi}^+_{cc} \Xi^+_{cc}, \, \overline{\Xi}^{++}_{cc}
\Xi^{++}_{cc}) ,
\end{equation}
and an analogous ${\bf M}^T_{\bar 3 3}$ for spin-3/2 states, and
\begin{equation}
{\bf d}^T_{\bar 3 3}  =  (d^1_0, \, d^8_0, \, d^8_1) .
\end{equation}

	Using the notation that the SU(3) chiral coefficients for the
spin-1/2 and spin-3/2 sextet are $a^R_I$ and $a^{*R}_I$ respectively,
the total $J=0$ (mass) chiral coefficients of SU(2) $\times$ SU(6) are
given, with the SU(6) rep in the lower index and SU(3) and isospin in
the upper indices, by
\begin{equation}
\begin{array}{cccc}
\left( \begin{array}{c}
X^{1,0}_1 \\ X^{1,0}_{405} \\ Y^{1,0}_{35}
\end{array} \right) &
= & \left( \begin{array}{ccc}
+\f{2}{\s{7}}  & +\s{\f{2}{7}}  & +\f{1}{\s{7}} \\
+\s{\f{2}{21}} & +\f{1}{\s{21}} & -\s{\f{6}{7}} \\
+\f{1}{\s{3}}  & -\s{\f{2}{3}}  &              0
\end{array} \right) & \left( \begin{array}{c}
a^{*1}_0 \\ a^1_0 \\ c^1_0
\end{array} \right)
\end{array} ,
\end{equation}
\begin{equation}
\begin{array}{cccc}
\left( \begin{array}{c}
X^{8,I}_{35} \\ X^{8,I}_{405} \\ Y^{8,I}_{35} \\ Y^{8,I}_{405}
\end{array} \right) & = & \left( \begin{array}{cccc}
+\f{1}{2} \s{\f{5}{2}} & +\f{\s{5}}{4}  & 0 & +\f{1}{4}      \\
+\f{1}{2\s{6}}         & +\f{1}{4\s{3}} & 0 & -\f{\s{15}}{4} \\
+\f{1}{\s{3}}          & -\s{\f{2}{3}}  & 0 & 0              \\
0                      & 0              & 1 & 0
\end{array} \right) & \left( \begin{array}{c}
a^{*8}_I \\ a^8_I \\ \f{1}{\s{2}} (b^8_I + \bar b^8_I) \\ c^8_I
\end{array} \right)
\end{array} ,
\end{equation}
\begin{equation}
Y^{10 + \overline{10}, 1}_{405} = +\f{1}{\s{2}} (b^{\overline{10}}_1
+ \bar b^{10}_1) ,
\end{equation}
\begin{equation}
\begin{array}{cccc}
\left( \begin{array}{c}
X^{27,I}_{405} \\ Y^{27,I}_{405}
\end{array} \right) & = & \left( \begin{array}{cc}
+\s{\f{2}{3}} & +\f{1}{\s{3}} \\ +\f{1}{\s{3}} & -\s{\f{2}{3}}
\end{array} \right) & \left( \begin{array}{c}
a^{* 27}_I \\ a^{27}_I
\end{array} \right)
\end{array} .
\end{equation}
$X$ and $Y$ are defined as chiral coefficients of SU(2) $\times$ SU(6)
in which the heavy-quark bilinear, as well as the light-quark
bilinear, has $J=0$ or 1, respectively; these can be thought of as
sets of bilinears that preserve or flip the heavy quark spin, and so
the latter are suppressed operators in the infinite mass limit.  In
adding chiral coefficients to their conjugates ({\it e.g.}
$\f{1}{\s{2}} (b^8_I + \bar b^8_I)$) we are imposing time-reversal
invariance (TRI) of the strong and electromagnetic Lagrangian, which
is responsible for the baryon masses.  Hermiticity alone would permit
combinations like $\f{i}{\s{2}} (b^8_I - \bar b^8_I)$, but these
violate TRI.

	For the states with two heavy quarks,
\begin{equation}
X^{R,I}_N = +\f{1}{\s{3}} (\s{2} d^{*R}_I + d^R_I) , \hspace{2em}
Y^{R,I}_N = +\f{1}{\s{3}} (d^{*R}_I - \s{2} d^R_I) .
\end{equation}
Presenting the Clebsch--Gordan coefficients in this factorized way is
equivalent to generating the isoscalar factors of SU(6) $\supset$
SU(3) $\times$ SU(2) for the product $\overline{\bf 21} \otimes {\bf
21}$ for $Q=1$ states, or $\bar {\bf 6} \otimes {\bf 6}$ for $Q=2$
states.  The analogous statement in (I) is that the isoscalar factors
for $\overline{\bf 56} \otimes {\bf 56}$ were generated.

	Finally, we extract the baryon mass combinations that form the
coefficients of operators transforming under particular reps of SU(2)
$\times$ SU(6).  Starting with the chiral coefficients $X$ and $Y$, we
read off the numerical coefficient of each baryon bilinear.  To obtain
the correct numerical coefficient in the corresponding mass
combination, we use the Wigner--Eckart theorem to remove the relevant
spin-SU(2) Clebsch--Gordan coefficient included in creating a total
$J=0$ bilinear, which is 1/2 for spin-3/2, 1/$\sqrt{2}$ for spin-1/2;
in short, one multiplies each spin-3/2 coefficient by $\s{2}$.  The
mass combination is denoted with a lower-case letter.  In the
following expressions, baryon masses are designated with their
symbols, and $\gamma$, $\delta_+$, and $\delta_0$ denote the
coefficients of $(\overline{\Sigma}^+_c \Lambda^+_c + h.c.)$,
$(\overline{\Xi}^{\prime +}_c \Xi^+_c + h.c.)$, and
$(\overline{\Xi}^{\prime 0}_c \Xi^0_c + h.c.)$, respectively.  Baryon
masses without isospin labels denote an average over all isospin
channels.
\begin{eqnarray} \label{q1rel}
x^{1,0}_1 & = & 2 (3\Sigma^*_c + 2\Xi^*_c + \Omega^*_c) + (3\Sigma_c +
2\Xi^\prime_c + \Omega_c) + (\Lambda_c + 2\Xi_c) , \nonumber \\
x^{1,0}_{405} & = & 2 (3\Sigma^*_c + 2\Xi^*_c + \Omega^*_c) +
(3\Sigma_c + 2\Xi^\prime_c + \Omega_c) -6(\Lambda_c + 2\Xi_c) ,
\nonumber \\
y^{1,0}_{35} & = & 3(\Sigma^*_c - \Sigma_c) + 2(\Xi^*_c -
\Xi^\prime_c) + (\Omega^*_c - \Omega_c) , \nonumber \\
x^{8,0}_{35} & = & 2(3\Sigma^*_c - \Xi^*_c -2\Omega^*_c) + (3\Sigma_c -
\Xi^\prime_c -2\Omega_c) + (\Lambda_c - \Xi_c) , \nonumber \\
x^{8,1}_{35} & = & 2[2(\Sigma^{*++}_c - \Sigma^{*0}_c) + (\Xi^{*+}_c -
\Xi^{*0}_c)] + [2(\Sigma^{++}_c - \Sigma^0_c) + (\Xi^{\prime +}_c -
\Xi^{\prime 0}_c)] + (\Xi^+_c - \Xi^0_c) , \nonumber \\
x^{8,0}_{405} & = & 2(3\Sigma^*_c - \Xi^*_c - 2\Omega^*_c) +
(3\Sigma_c - \Xi^\prime_c - 2\Omega_c) - 15(\Lambda_c - \Xi_c) ,
\nonumber \\
x^{8,1}_{405} & = & 2[2(\Sigma^{*++}_c - \Sigma^{*0}_c) + (\Xi^{*+}_c
- \Xi^{*0}_c)] + [2(\Sigma^{++}_c - \Sigma^0_c) + (\Xi^{\prime +}_c -
\Xi^{\prime 0}_c)] - 15(\Xi^+_c - \Xi^0_c) , \nonumber \\
y^{8,0}_{35} & = & 3(\Sigma^*_c - \Sigma_c) - (\Xi^*_c - \Xi^\prime_c)
- 2(\Omega^*_c - \Omega_c) , \nonumber \\
y^{8,1}_{35} & = & 2[(\Sigma^{*++}_c - \Sigma^{*0}_c) - (\Sigma^{++}_c
- \Sigma^0_c)] + [(\Xi^{*+}_c - \Xi^{*0}_c) - (\Xi^{\prime +}_c -
\Xi^{\prime 0}_c)] , \nonumber \\
y^{8,0}_{405} & = & \delta_+ + \delta_0 , \nonumber \\
y^{10+\overline{10},1}_{405} & = & \gamma - \delta_+ + \delta_0 ,
\nonumber \\
y^{8,1}_{405} & = & 2\gamma + \delta_+ - \delta_0 , \nonumber \\
x^{27,0}_{405} & = & 2(\Sigma^*_c - 2\Xi^*_c + \Omega^*_c) + (\Sigma_c
-2\Xi^\prime_c + \Omega_c) , \nonumber \\
x^{27,1}_{405} & = & 2[(\Sigma^{*++}_c - \Sigma^{*0}_c) - 2(\Xi^{*+}_c
- \Xi^{*0}_c)] + [(\Sigma^{++}_c - \Sigma^0_c) - 2(\Xi^{\prime +}_c -
\Xi^{\prime 0}_c)] , \nonumber \\
x^{27,2}_{405} & = & 2(\Sigma^{*++}_c - 2\Sigma^{*+}_c +
\Sigma^{*0}_c) + (\Sigma^{++}_c - 2\Sigma^+_c + \Sigma^0_c) ,
\nonumber \\
y^{27,0}_{405} & = & (\Sigma^*_c - \Sigma_c) - 2(\Xi^*_c -
\Xi^\prime_c) + (\Omega^*_c - \Omega_c) , \nonumber \\
y^{27,1}_{405} & = & [(\Sigma^{*++}_c - \Sigma^{*0}_c) -
(\Sigma^{++}_c - \Sigma^0_c)] - 2[(\Xi^{*+}_c - \Xi^{*0}_c) -
(\Xi^{\prime +}_c - \Xi^{\prime 0}_c)] , \nonumber \\
y^{27,2}_{405} & = & (\Sigma^{*++}_c - 2\Sigma^{*+}_c +
\Sigma^{*0}_c) - (\Sigma^{++}_c - 2\Sigma^+_c + \Sigma^0_c) .
%
\end{eqnarray}
Likewise, for the doubly-charmed baryons, we obtain
\vspace{-1em}
\begin{eqnarray} \label{q2rel}
x^{1,0}_1 & = & 2(\Omega^*_{cc} + 2\Xi^*_{cc}) + (\Omega_{cc} +
2\Xi_{cc}) , \nonumber \\
x^{8,0}_{35} & = & 2(\Omega^*_{cc} - \Xi^*_{cc}) + (\Omega_{cc} -
\Xi_{cc}) , \nonumber \\
x^{8,1}_{35} & = & 2(\Xi^{*+}_{cc} - \Xi^{*++}_{cc}) + (\Xi^+_{cc} -
\Xi^{++}_{cc}) , \nonumber \\
y^{1,0}_{35} & = & (\Omega^*_{cc} - \Omega_{cc}) + 2(\Xi^*_{cc} -
\Xi_{cc}) , \nonumber \\
y^{8,0}_{35} & = & (\Omega^*_{cc} - \Omega_{cc}) - (\Xi^*_{cc} -
\Xi_{cc}) , \nonumber \\
y^{8,1}_{35} & = & (\Xi^{*+}_{cc} - \Xi^{*++}_{cc}) - (\Xi^+_{cc} -
\Xi^{++}_{cc}) .
\end{eqnarray} 
\begin{table}
\caption{Current experimental values for the heavy baryon masses.}
\label{mass}
\begin{tabular}{lll}
State & Mass (MeV) & Ref. \\
\hline\hline
$\Lambda_c$        & 2285.1 $\pm$ 0.6 & \cite{PDG} \\
$\Sigma^{++}_c$    & 2453.1 $\pm$ 0.6 & \cite{PDG} \\
$\Sigma^+_c$       & 2453.8 $\pm$ 0.9 & \cite{PDG} \\
$\Sigma^0_c$       & 2452.4 $\pm$ 0.7 & \cite{PDG} \\
$\Xi^+_c$          & 2465.1 $\pm$ 1.6 & \cite{PDG} \\
$\Xi^0_c$          & 2470.3 $\pm$ 1.8 & \cite{PDG} \\
$\Xi^{\prime +}_c$ & 2560 $\approx \pm$ 15      & \cite{WA89} \\
$\Omega^0_c$       & 2699.9 $\pm$ 1.5 $\pm$ 2.5 & \cite{fnalom} \\
$\Sigma^{*++}_c$   & 2530 $\pm$ 5 $\pm 5$       & \cite{SKAT} \\
$\Xi^{*+}_c$       & 2644.6 $\pm$ 2.3 & \cite{xicleo} \\
$\Xi^{*0}_c$       & 2642.8 $\pm$ 2.2 & \cite{xicleo} \\
\hline
$\Lambda_b$       & 5641 $\pm$ 50         & \cite{PDG}    \\
                  & 5623 $\pm$ 5  $\pm$ 4 & \cite{cdflam} \\
                  & 5614 $\pm$ 21 $\pm$ 4 & \cite{allam}  \\
                  & 5668 $\pm$ 16 $\pm$ 8 & \cite{dellam} \\
$\Sigma^\pm_b$    & 5841 $\pm$ 16 $\pm$ 8 & \cite{delsig} \\
$\Sigma^{*\pm}_b$ & 5897 $\pm$ 16 $\pm$ 8 & \cite{delsig} \\
\end{tabular}
\end{table}
\begin{table}
\caption{Estimated magnitudes of mass difference combinations.}
\label{comb}
\begin{tabular}{lll|lll}
Combination & Est.\ mag.\ & (MeV) \, &
Combination & Est.\ mag.\ & (MeV) \, \\
\hline\hline
\multicolumn{6}{c}{$Q=1$} \\
\hline
$x^{1,0}_{405}$  & 9$\Lambda_{\chi} \delta$ & 2700 &
$y^{10 + \overline{10},1}_{405}$ & 3$\Lambda_{\chi} \theta \epsilon^2
\epsilon^\prime$ & 0.27 \\
$y^{1,0}_{35}$   & 3$\Lambda_{\chi} \theta$ & 600 &
$y^{8,1}_{405}$  & 4$\Lambda_{\chi} \theta \epsilon^\prime$ & 4 \\
$x^{8,0}_{35}$   & 5$\Lambda_{\chi} \epsilon$   & 1500 &
$x^{27,0}_{405}$ & 3$\Lambda_{\chi} \epsilon^2$ & 270 \\
$x^{8,1}_{35}$   & 5$\Lambda_{\chi} \epsilon^\prime$ & 25 &
$x^{27,1}_{405}$ & $\f{9}{2}\Lambda_{\chi} \epsilon \epsilon^\prime$
& 6.8 \\
$x^{8,0}_{405}$  & 12$\Lambda_{\chi} \delta \epsilon$ & 1080 &
$x^{27,2}_{405}$ & 3$\Lambda_{\chi} \epsilon^{\prime\prime}$ & 3 \\
$x^{8,1}_{405}$  & 12$\Lambda_{\chi} \delta \epsilon^\prime$ & 18 &
$y^{27,0}_{405}$ & 2$\Lambda_{\chi} \theta \epsilon^2$ & 36 \\
$y^{8,0}_{35}$   & 3$\Lambda_{\chi} \theta \epsilon$ & 180 &
$y^{27,1}_{405}$ & 3$\Lambda_{\chi} \theta \epsilon \epsilon^\prime$ &
0.9 \\
$y^{8,1}_{35}$   & 3$\Lambda_{\chi} \theta \epsilon^\prime$ & 3 &
$y^{27,2}_{405}$ & 2$\Lambda_{\chi} \theta \epsilon^{\prime\prime}$ &
0.4 \\
$y^{8,0}_{405}$  & 2$\Lambda_{\chi} \theta \epsilon$ & 120 &
& & \\
\hline
\multicolumn{6}{c}{$Q=2$} \\
\hline
$x^{8,0}_{35}$ & $\f{3}{2} \Lambda_{\chi} \epsilon$ & 450 &
$y^{8,0}_{35}$ & $\Lambda_{\chi} \theta \epsilon$   & 60 \\
$x^{8,1}_{35}$ & $\f{3}{2} \Lambda_{\chi} \epsilon^\prime$ & 7.5 &
$y^{8,1}_{35}$ & $\Lambda_{\chi} \theta \epsilon^\prime$ & 1 \\
$y^{1,0}_{35}$ & $\f{3}{2} \Lambda_{\chi} \theta$ & 300 
\end{tabular}
\end{table}
\begin{table}
\caption{Mass combinations versus experimental values.}
\label{compare}
\begin{tabular}{lr|rrr}
Masses & Exp.\ (MeV) & \multicolumn{1}{c}{Combination} & Est.\ mag.\ &
(MeV) \\
\hline\hline
$\f{1}{21} (6\Sigma^*_c \! + \! 6\Xi^*_c \! + \! 3\Sigma_c \! + \!
3\Omega_c \! + \! \Lambda_c \!  + \! 2\Xi_c)$ & $2558 \pm 2$ &
$\f{1}{105} (5x^{1,0}_1 + 2y^{8,0}_{35} - 6y^{27,0}_{405})$ &
${\rm M}_c \pm \f{2}{35} \Lambda_\chi \theta \epsilon$
& ${\rm M}_c \pm 3$ \\
$2\Sigma^*_c + 2\Xi^*_c + \Sigma_c + \Omega_c - 2\Lambda_c - 4\Xi_c$ &
$1059 \pm 16$ &
$\f{1}{15}(5x^{1,0}_{405} + 2y^{8,0}_{35} - 6y^{27,0}_{405})$ &
$3\Lambda_\chi \delta$ & 900 \\
$\Sigma^*_c - \Sigma_c$ & $77 \pm 7$ &
$\f{1}{30} (5y^{1,0}_{35} + 4y^{8,0}_{35} + 3y^{27,0}_{405})$ &
$\f{1}{2} \Lambda_\chi \theta$ & 100 \\
$3\Sigma^*_c - 3\Xi^*_c + 6\Sigma_c - 6\Omega_c + \Lambda_c - \Xi_c$ &
$-2000 \pm 30$ & $\f{1}{5}(5x^{8,0}_{35} - 7y^{8,0}_{35} +
6y^{27,0}_{405})$ &
$5\Lambda_\chi \epsilon$ & 1500 \\
$\Sigma^*_c - \Xi^*_c + 2\Sigma_c - 2\Omega_c - 5\Lambda_c + 5\Xi_c$ &
$306 \pm 12$ & $\f{1}{15}(5x^{8,0}_{405} - 7y^{8,0}_{35} +
6y^{27,0}_{405})$ &
$4\Lambda_\chi \delta \epsilon$ & 360 \\
$2\Sigma^*_c - 2\Xi^*_c - \Sigma_c + \Omega_c$ & $19 \pm 15$ &
$\f{1}{15}(5x^{27,0}_{405} + 6y^{8,0}_{35} + 2y^{27,0}_{405})$ &
$\f{3}{2} \Lambda_\chi \epsilon^2$ & 135 \\
\hline
$3\Xi^*_{c1} + 6\Sigma_{c1} + \Xi_{c1}$ & $2.9 \pm 8.5$ &
$\f{1}{15}(5x^{8,1}_{35} -7y^{8,1}_{35} - 6y^{27,1}_{405})$ &
$5\Lambda_\chi \epsilon^\prime$ & 25 \\
$\Xi^*_{c1} + 2\Sigma_{c1} - 5\Xi_{c1}$ & $29 \pm 11$ &
$\f{1}{15}(5x^{8,1}_{405} - 7y^{8,1}_{35} - 6y^{27,1}_{405})$ &
$4\Lambda_\chi \delta \epsilon^\prime$ & 6 \\
$\Sigma_{c1} - 2\Xi^*_{c1}$ & $-1.9 \pm 5.2$ &
$\f{1}{15}(5x^{27,1}_{405} - 6y^{8,1}_{35} + 2y^{27,1}_{405})$ &
$\f{3}{2}\Lambda_\chi \epsilon \epsilon^\prime$ & 2.3 \\
\hline
$\Sigma^{++}_c - 2\Sigma^+_c + \Sigma^0_c$ & $-2.1 \pm 1.3$ &
$\f{1}{3}(x^{27,2}_{405} - 2y^{27,2}_{405})$ &
$\Lambda_\chi \epsilon^{\prime\prime}$ & 1
\end{tabular}
\end{table}

\begin{thebibliography}{99}
\bibitem{phase}This is the phase convention for conjugate
reps that gives the fundamental quark triplet as $(u,d,s)$, but the
fundamental antitriplet as $(\bar u, -\bar d, \bar s)$; it guarantees
that all raising and lowering operators produce positive coefficients.
Because mass operators are quadratic bilinears, in our case it has no
effect except to redefine the sign of $\Xi^+_c$ mixing terms.
\end{thebibliography}
\end{document}